\documentclass[preprint,amssymb,
 aps]{revtex4-1}
\usepackage[dvips]{graphicx}

\begin{document}

\title{Lewis-Riesenfeld quantization and  $SU(1,1)$ coherent states for 2D damped  harmonic oscillator}

\date{\today}

\author{Lat\'evi M. Lawson}
\email{ latevi.lawson@imsp-uac.org; lawmenx@gmail.com}

\author{Gabriel Y. H. Avossevou }
\email{gabriel.avossevou@imsp-uac.org}

\affiliation{Laboratoire de Recherche en Physique Th\'eorique (URPT),\\
 Institut de Math\'ematiques et de Sciences  Physiques (IMSP)\\
 Universit\'e d'Abomey-Calavi (UAC)\\
01 BP 613 Porto-Novo, Rep du B\'enin.}

\author{ Laure Gouba}
\email{lgouba@ictp.it} 
\affiliation{The Abdus Salam International Centre for Theoretical Physics (ICTP),\\
             Strada Costiera 11, I-34151 Trieste, Italy}   
             
\begin{abstract}
In this paper we study a two-dimensional [2D] rotationally symmetric harmonic oscillator with 
time-dependent frictional force. At the classical level, we solve the  equations of  motion  for a particular case 
of the  time-dependent coefficient of  friction. 
At the quantum level, we use  the Lewis-Riesenfeld procedure of  invariants   to construct
exact solutions  for the corresponding time-dependent Schr\"{o}dinger equations.
The eigenfunctions obtained are  in terms of the  generalized Laguerre polynomials.  By mean  of the solutions 
we verify a generalization  version of the Heisenberg's uncertainty relation  and  derive the generators of the $su(1,1)$ Lie algebra.
Based on these generators, we construct the coherent states $\grave{\textrm{a}}$ la  
Barut-Girardello and $\grave{\textrm{a}}$ la  Perelomov 
and  respectively study their properties.
\end{abstract}  

\maketitle

\section{Introduction}\label{Intro}
\label{intro}
The one-dimensional [$1$D] harmonic oscillator  is one of the most simplest and  fundamental classical
as well as quantum system studied in the literature. 
However, the study of the two-dimensional  [$2$D] harmonic oscillator in quantum mechanics  for the case of 
 the rotationally symmetric oscillator turns out to be  interesting and less explored. 
In fact, it is  more difficult   to solve when the problem  involves  time-dependent parameters.
 
In the last few decades the problem of the time-dependent  quantum systems has received 
a great interest since Lewis and Riesenfeld have introduced an  excellent method of invariants to solve the time-dependent 
Schr\"{o}dinger  equation \cite{1}. 
This method stimulated  some interest in using the invariants for solving $1$D and $2$D time-dependent harmonic oscillators problems
\cite{2,3,4,5,6,7,8,9,10,11,12,13,14,15,16,16',17,18,19,20,21,22}.
The  $1$D damped harmonic oscillator has  been  extensively studied in the literature \cite{23,24,25,26,27}, while its generalization in 
two-dimensions as far as we know is less explored.  
We  discuss a system of two-non-interacting damped oscillators with equal time-dependent coefficients of friction and equal time-dependent
frequencies.
  
In section \ref{sec2}, we study the system at the classical level and  formulate the corresponding quantum system. We solve  the classical  equations
of motion for  a constant coefficient of friction and for some  particular cases of frequencies.
   
In section \ref{sec3}, we use the Lewis-Riesenfeld's method to  construct the invariant operator $\hat I(t)$. The eigenvalues and the eigenfunctions
 of  the invariants  are  calculated explicitly by operators methods, the key  element being the introduction of  an appropriate unitary
operator. We  derive then   a  conserved  angular momentum $\hat L_z$ that  is simultaneously commuting with the invariant operator
$\hat I(t)$ and  the Hamiltonian $\hat H(t)$. However, the three operators cannot be simultaneously diagonalized at this stage 
of the problem.
   
In section \ref{sec4}, we introduce  the helicity Fock basis in order  to simultaneously diagonalize the operators 
$\hat I(t), \hat H(t)$ and $\hat L_z$. The rotationally symmetry of the system has been useful in  determining an orthogonal
basis of the Hilbert space  for the procedure of the  simultaneous diagonalization. Then we derive the exact solution of the Schr\"{o}dinger
equations in terms of generalized Laguerre polynomials.
   
 In section \ref{sec5}, we use the eigenfunctions of the Hamiltonian to verify a generalization version of the Heisenberg's uncertainty relations that
 are formulated  following the standard arguments  as follows: for the simultaneous measurement  of two observables $\hat A$ and $\hat B$ in the 
 states $|\psi\rangle$, the uncertainty satisfy the inequality
 \begin{equation}\label{He}
 \Delta \hat A  \Delta \hat B\geq\frac{\hbar}{2}\big |\langle\psi|[\hat A,\hat B]|\psi\rangle\big|,\,
\end{equation}
 where $ \Delta \hat A  $ and $ \Delta\hat  B  $  are respectively the dispersions  defined as
 \begin{eqnarray} \label{v3}
  \Delta \hat A=\sqrt{\langle\psi|\hat A^2|\psi\rangle-\langle\psi|\hat A|\psi\rangle^2},\,\,
   \Delta \hat B=\sqrt{\langle\psi|\hat B^2|\psi\rangle-\langle\psi|\hat B|\psi\rangle^2}.
 \end{eqnarray}
  Similar discussions can be also  read in \cite{10}.
  
In section \ref{sec6}, we  derive  from the  solution of  the system the hidden  generators of the $su(1,1)$ Lie  algebra.
We proceed by  the factorization method as developped in \cite{28,29} to find  the hidden symmetry of the system and  derive 
from the eigenfunctions the related raising and lowering operators which generate the $su(1,1)$ Lie algebra.
  
In section \ref{sec7}, we discuss the $SU(1,1)$ coherent  states $\grave{\textrm{a}}$ la Barut-Girardello \cite{30} and $\grave{\textrm{a}}$ la Perelomov \cite{31}.
A brief  story  about  these   coherent states  is that in $ 1926$'s
Schr\"{o}dinger introduced for the first time in quantum mechanics the  semiclassical states defined 
 as the minimum uncertainty Gaussian states whose dynamics has
maximum similarity to classical oscillator  \cite{32}. These states were rediscovered by 
Glauber in the framework of quantum optics in early  $1960'$s  \cite{33}. They are defined as eigenstates of the annihilation
operator of harmonic oscillator and were obtained by action of the Weyl-Heisenberg operator on the ground state.
These  coherent states introduced by Glauber have inspired respectively Barut-Girardello \cite{30} 
and Perelomov \cite{31} in  constructing the coherent states
for $SU (1, 1)$ Lie algebraic group through different approaches. The Barut-Girardello  and the Perelomov coherent states  
gained  lot  of applications,  for instance in the fields of quantum  optics \cite{34,35}, quantum computation \cite{36,37}
and quantum mechanics \cite{38,39,40}.

The conclusion is given in section \ref{sec8}.

\section{The Model}\label{sec2}
We consider in two-dimensional configuration space, the system of two non-interacting damped oscillators with equal time-dependent
 coefficients of friction and equal time-\\dependent frequencies. The equations of motion are given by
  \begin{equation}\label{z}
  \left\{
  \begin{array}{rcr}
   \ddot{x}_1 +\eta(t)\dot{x}_1+\omega^2(t) x_1=0, \\
   \ddot{x}_2 +\eta(t)\dot{x}_2+\omega^2(t) x_2=0,
   \end{array}
\right.
  \end{equation}
where $\eta(t)$ is the time-dependent  coefficient of friction, $\omega(t)$ is the time-dependent frequency and the dot represents
time-derivative.\\
These equations of motion  may be derived from the Lagrangian 
\begin{equation}\label{eq1}
 L(x_1,x_2,\dot{x}_1,\dot{x}_2,t)=f^{-1}(t)\left[\frac{m}{2}(\dot{x}_1^2+ \dot{x}_2^2) -\frac{m\omega^2(t)}{2}(x_1^2+x_2^2)\right],
\end{equation}
where $f$ is an arbitrary function such  that $f(t)=e^{-\int_0^t\eta(t')dt'}$ or  $\eta(t)=-\frac{d}{dt}[\ln f(t)]$.\\
Let consider $R(\vartheta)$,  the rotation matrix in the plane  which transforms  coordinates $x (x_1,x_2)$ into others $x'(x_1',x_2')$  such  as
\begin{eqnarray}
 x'=R(\vartheta)x,\,\,\,\,
 R(\vartheta)= 
   \left (
   \begin{array}{cc}
 \cos\vartheta&-\sin \vartheta\\\sin\vartheta &\cos \vartheta
 \end{array}
   \right ), \,\,\,\, \,\,\,\vartheta\in \mathbb R.
\end{eqnarray}
This  transformation  preserves the invariance of  
the Lagrangian. This rotational invariance   in the plane manifests the presence of 
the Noether  charge  which correspond to the angular-momentum of the system.

The canonical momentum associated with the variables $x_1$ and $x_2$ are
\begin{equation}\label{w}
   \left\{
  \begin{array}{rcr}
   p_1=\frac{\partial L}{\partial \dot{x}_1}=f^{-1}(t) m \dot{x}_1, \\
   p_2=\frac{\partial L}{\partial \dot{x}_2}=f^{-1}(t) m \dot{x}_2.
    \end{array}
\right.
  \end{equation}
  The  Hamiltonian is given by
 \begin{eqnarray} \label{x1}
  H(x_1,x_2,p_1,p_2,t)&=& \dot{x}_1 p_1+\dot{x}_2 p_2-L\cr
  &=&\frac{f(t)}{2m}\left(p_1^2+p_2^2\right)+ f^{-1}(t)\frac{ m\omega^2(t)}{2} \left (x_1^2+x_2^2\right).
 \end{eqnarray}
 We recover the $2$D  Hamiltonian that describes  the dissipative system  previously introduced in one dimension 
 by Pedrosa \cite{26,27}.
 For $f(t)=1$ and $ f(t)=\exp\left(-\gamma t\right)$ with $\omega(t)=\omega_0$ where  $\gamma,\,\omega_0$ are positive constants,  the Hamiltonian
 (\ref{x1}) is respectively reduced to the ordinary $2$D  harmonic oscillator and   the $2$D Caldirola and Kanai Hamiltonian \cite{41,42}.
 
 Since we are in two dimensional configuration space, we can look for the solutions of the classical equations  in the complex system by
 setting $z=x_2+ix_1$. The classical  equation of motion  in term of the coordinate $z$  is 
\begin{equation}\label{x2}
  \ddot{z} +\eta(t) \dot{z}+\omega^2(t) z=0.
 \end{equation}
For $\eta(t)=\gamma$ and $\omega(t)=\omega_0$, the equation (\ref{x2}) takes the form
\begin{equation}
  \ddot{z} +\gamma \dot{z}+\omega_0^2 z=0,
 \end{equation}
 and the classical solutions  are \cite{43}
 \begin{eqnarray}
 z(t)=  \left\{
  \begin{array}{l}
  e^{-\frac{1}{2}\gamma t}\left[ A_1\exp\left(\frac{1}{2}\tau t\right)+
  A_2\exp\left(-\frac{1}{2}\tau t\right)\right] \,\,\mbox{if} \,\,\tau^2=\gamma^2-4\omega_0^2>0, \\
    e^{-\frac{1}{2}\gamma t}\left[ A_1\sin\left(\frac{1}{2}\tau t\right)+
  A_2\cos\left(\frac{1}{2}\tau t\right)\right]\,\,\,\mbox{if} \,\,\,\tau^2=4\omega_0^2-\gamma^2>0,\\
  e^{-\frac{1}{2}\gamma t} (A_1+  A_2) \,\,\,\mbox{if} \,\,\gamma^2=4\omega_0^2,
  \end{array}
\right.
  \end{eqnarray}
where $ A_1$ and $A_2$ are constants.

 For $\omega(t)= \omega_0 e^{-\frac{1}{2}\gamma t}$, the  equation can be rewritten as follows 
 \begin{equation}
  \ddot{z} +\gamma \dot{z}+\omega_0^2 e^{-\gamma t} z=0.
 \end{equation}
 The solution is given by \cite{24,43} 
  \begin{equation}
   z(t)=  e^{-\frac{1}{2}\gamma t}\left [ B_1 J_1\left(\frac{2\omega_0}{\gamma}e^{-\frac{1}{2}\gamma t}\right)+ B_2Y_1\left(
   \frac{2\omega_0}{\gamma}e^{-\frac{1}{2}\gamma t}\right)\right],
  \end{equation}
where $J_k$ and $Y_k$ are respectively Bessel functions of first and second kind, $B_1$ and $B_2$ are constants.

For  $\omega(t)= \omega_0 e^{-\gamma t}$,
the solution is known  to be \cite{24,43}
\begin{equation}
 z(t)= C_1 \cos\left(\frac{\omega_0e^{-\gamma t}}{\gamma}\right)+ C_2\sin\left(\frac{\omega_0 e^{-\gamma t}}{\gamma}\right),
\end{equation}
where $C_1$ and $C_2$ are constants.

At the quantum level, the corresponding Hamiltonian operator describing the system reads
\begin{eqnarray}\label{H1}
 \hat H(t) = \frac{f(t)}{2m}\left(\hat p_1^2+\hat p_2^2\right)+ f^{-1}(t)\frac{ m\omega^2(t)}{2} \left (\hat x_1^2+\hat x_2^2\right),
\end{eqnarray}
where the position operators $\hat x_1, \hat x_2$ and the momentum operators $\hat p_1,\hat p_2$ satisfy the canonical commutation relations
\begin{eqnarray}
 [\hat x_i,\hat p_j]=i\hbar\mathbf{I}\delta_{ij},\,\,\,\,\,[\hat x_i,\hat x_j]=0=[\hat p_i,\hat p_j],\,\,\, i,j=1,2.
\end{eqnarray}
 To  diagonalize  this Hamiltonian, many methods can be considered to achieve this end \cite{1,6,44,45,46,47,48}.
 Among them, we have the Lewis-Riesenfield method  based on the construction of the Hermitian invariant operator \cite{1}.
 
 \section{Construction and eigensystems of the  invariant operator}\label{sec3}

 To construct the  exact invariant operator for the quantum system described by the time-dependent Hamiltonian  (\ref{H1}), 
 we use the dynamic invariant  method formulated  by Lewis and Riesenfeld \cite{1}.  
 
 Now, we look for the invariant in the form
\begin{equation}\label{x4}
 \hat I(t)=\alpha (t)\hat J_++\beta(t)\hat J_- +\delta (t) \hat J_0,
\end{equation}
where $\alpha,\beta,\delta$ are time-dependent real coefficients  and  
$\hat J_+=\frac{1}{2}(\hat x_1^2+\hat x_2^2)$, $\hat J_-=\frac{1}{2}(\hat p_1^2+\hat p_2^2)$, 
$\hat J_0=\frac{1}{2}(\hat x_1\hat p_1+\hat p_1\hat x_1+\hat x_2\hat p_2+\hat p_2\hat x_2)$ satisfy the following commutation relations
\begin{eqnarray}
 [\hat J_+,\hat J_-]=i\hat J_0; \,\, [\hat J_0,\hat J_\pm]=\pm2i\hat J_\pm.
\end{eqnarray}
The Hamiltonian (\ref{H1}) is rewritten in term  of the latter operators as follows
\begin{equation}
 \hat H(t)=\frac{f(t)}{m}\hat J_-+ f^{-1}(t) m\omega^2(t) \hat J_+.
\end{equation}
 To determine  an explicit form  of the Hermitian invariant (\ref{x4}), one solves  
 the  following equation
 \begin{equation}\label{C1}
  \frac{d\hat I (t)}{dt}=\frac{\partial \hat I(t)}{\partial t}+\frac{1}{i}[\hat I(t),\hat H(t)]\equiv0,
  \end{equation}
  where  $\hbar=1$. By expansion of equation (\ref{C1}),
we obtain the  first-order linear differential equations for the unknown coefficient functions
\begin{eqnarray}
 \dot{\alpha}-2f^{-1}m\omega^2\delta=0,\\
 \dot{\beta}+\frac{2f}{m}\delta=0,\\
 \dot{\delta}+\frac{f}{m}\alpha-f^{-1}m\omega^2\beta=0.
\end{eqnarray}
As in \cite{1,14},
it  is convenient to introduce another real function $\rho (t)$  
\begin{equation}\label{x5}
 \beta(t)=\rho^2(t).
\end{equation}
For an arbitrary positive constant $\nu$, the other coefficients are 
\begin{eqnarray}\label{x6}
 \delta (t)=-mf^{-1}\dot{\rho}\rho,\,\,\,\,\,\alpha(t)=\frac{\nu^2}{\rho^2}+m^2f^{-2}{\dot{\rho}}^2.
\end{eqnarray}
Replacing (\ref{x5}), (\ref{x6}) in (\ref{x4}), the Hermitian invariant acquires the form
\begin{eqnarray}
 \hat I(t)=\frac{1}{2}\left[\left(mf^{-1}\dot{\rho}\hat x_1-\rho \hat p_1\right)^2 +\frac{\nu^2}{\rho^2}\hat x_1^2+
 \left(mf^{-1}\dot{\rho}\hat x_2-\rho \hat p_2\right)^2 +\frac{\nu^2}{\rho^2}\hat x_2^2\right],
\end{eqnarray}
where the function $\rho$ is the solution of the so-called  Ermakov-Pinney equation \cite{49}
\begin{equation}\label{e4}
 \ddot{\rho}+\eta \dot{\rho}+\omega^2\rho=\frac{\nu^2 f^2}{m^2\rho^3}.
\end{equation}

Next we  determine the spectrum of  the invariant operator by solving the eigenvalue equation
\begin{equation}\label{I1}
 \hat I(t)\phi( x_1, x_2,t)=E \phi( x_1, x_2,t),
 \end{equation}
 where $E$ is a constant, $\phi(x_1,x_2,t)$ is element of Hilbert space   
$\mathcal{H}$  on which this operator is defined.

In order to solve  equation (\ref{I1})  we introduce the unitary operator 
that is written as follows
\begin{equation}
 \hat U=\exp\left[-\frac{imf^{-1}\dot{\rho}}{2\rho}(\hat x_1^2+\hat x_2^2)\right],\,\,\,\,\,  \hat U^\dag \hat U= \hat U\hat U^\dag =\mathbf{I}.
\end{equation}
Setting
\begin{equation}
 U\phi(x_1,x_2,t)=\phi'(x_1,x_2,t),
\end{equation}
 and
\begin{equation}
 \hat I'(t)=\hat U \hat I\hat U^\dag= \frac{1}{2}\left[\rho^2(\hat p_1^2+\hat p_2^2)+\frac{\kappa^2}{\rho^2}(\hat x_1^2+\hat x_2^2)\right],
\end{equation}
it is  easy to verify that
\begin{equation}\label{I2}
 \hat I'(t)\phi'( x_1, x_2,t)=E \phi'( x_1, x_2,t),
\end{equation}
where $\phi'( x_1, x_2,t)\in\mathcal{H}$.
To achieve the diagonalization of equation (\ref{I2}) as clear as possible, we introduce the lowering
and raising operators given by
\begin{eqnarray}
 a_1'&=&\frac{1}{\sqrt{2\nu}}\left(\frac{\nu}{\rho}\hat x_1+i\rho \hat p_1\right),\,\,\,\,
 {a'}_1^\dag=\frac{1}{\sqrt{2\nu}}\left(\frac{\nu}{\rho }\hat x_1-i\rho \hat p_1\right),\\
  a_2'&=&\frac{1}{\sqrt{2\nu}}\left(\frac{\nu}{\rho}\hat x_2+i\rho \hat p_2\right),\,\,\,\,
   {a'}_2^\dag=\frac{1}{\sqrt{2\nu}}\left(\frac{\nu}{\rho}\hat x_2-i\rho \hat p_2\right),
\end{eqnarray}
which satisfy the following commutation relations
\begin{eqnarray}
 [a_1',{a'}_1^\dag]=\mathbf{I}=  [a_2',{a'}_2^\dag],\,\,\,\,\,[a_1',a_2']=0=  [{a'}_1^\dag,{a'}_2^\dag].
\end{eqnarray}

Let us consider any  nonnegative integers $n_1,n_2$ and $|\phi'_{n_1,n_2}(t)\rangle$ the orthonormalized
Fock space  such as
\begin{eqnarray}
 |\phi'_{n_1,n_2}(t)\rangle&=&\frac{1}{\sqrt{n_1!n_2!}}\left({a'_1}^\dag\right)^{n_1} \left({a'_2}^\dag\right)^{n_2}|\phi'_{0,0}(t)\rangle,\label{to}\\
 \langle \phi'_{n_1,n_2}(t)|\phi'_{m_1,m_2}(t)\rangle&=&\delta_{n_1,m_1}\delta_{n_2,m_2},
\end{eqnarray}
with $|\phi'_{0,0}(t)\rangle$ is a normalized state annihilated  by $a'_1,a'_2$.

In order to determine the exact solution $\phi_{n_1,n_2}(x_1,x_2,t)$ of the invariant operator $I(t)$, we first express the  
 ground state $|\phi_{0,0}(t)\rangle$ in the configuration space basis as follows 
\begin{eqnarray}
  \phi_{0,0}(x_1,x_2,t)&=&U^\dag\langle x_1|\phi_0'(t)\rangle\langle x_2|\phi_0'(t)\rangle\cr
                        &=&\left(\frac{\nu}{\pi\rho^2}\right)^{\frac{1}{2}}\exp\left[ \left(imf^{-1}\frac{\dot{\rho}}{\rho}-\frac{\nu}{\rho^2}\right)
                       \left (\frac{x_1^2+x_2^2}{2}\right)\right].
 \end{eqnarray}
Then, the nth eigenfunction are obtained from (\ref{to}) as 
\begin{eqnarray}\label{ajou}
  \phi_{n_1,n_2}(x_1,x_2,t)&=& U^\dag \phi_{n_1,n_2}'(x_1,x_2,t)\cr
                           &=&\frac{1}{\rho}\left(\frac{\nu}{2^{n_1+n_2}\pi\, n_1!n_2!}\right)^{\frac{1}{2}}
                             H_{n_1}\left(x_1\frac{\sqrt{\nu}}{\rho}\right) 
                            H_{n_2}\left(x_2\frac{\sqrt{\nu}}{\rho}\right)\cr&&\times\exp\left[\left(imf^{-1}\frac{\dot{\rho}}{\rho}-\frac{\nu}{\rho^2}\right)\left(\frac{x_1^2}{2}
  +\frac{x_2^2}{2}\right)\right],
\end{eqnarray}
 where $H_{n_1}$ and $H_{n_2}$ are the Hermite polynomials of order $n_1$ and $n_2$.
 
 To obtain the eigenvalues $E_{n_1,n_2}$ of the invariant operator $\hat I(t)$, let us introduce a
 new pair of raising and lowing  operators define as
 \begin{eqnarray}
 a_j&=&U^\dag a'_j U=\frac{1}{\sqrt{2\nu}}\left(mf^{-1}\dot{\rho}\hat x_j-\rho \hat p_j+i\frac{\nu}{\rho}\hat x_j\right)\label{x7},\\
 a_j^\dag&=&U^\dag {a'}_j^\dag U=\frac{1}{\sqrt{2\nu}}\left(mf^{-1}\dot{\rho}\hat x_j-\rho \hat p_j-i\frac{\nu}{\rho}\hat x_j\right) \label{x8}.
\end{eqnarray}
 with $j=1,2$. In term of these operators the invariant  operator $\hat I(t)$ takes the form
 \begin{eqnarray}
 \hat I(t)&=& \nu\left(a_1^\dag a_1+a_2^\dag a_2+\mathbf{I}\right).
\end{eqnarray}
The action of $a_j$ and $a_j^\dag$ on $ |\phi_{n_j}(t)\rangle$  finds expression in 
\begin{eqnarray}
  a_j^\dag|\phi_{n_j}(t)\rangle&=&\sqrt{n_j+1}|\phi_{n_j+1}(t)\rangle,\\
  a_j|\phi_{n_j}(t)\rangle&=& \sqrt{n_j}|\phi_{n_j-1}(t)\rangle,\\
  a_j^\dag a_j|\phi_{n_j}(t)\rangle&=& n_j|\phi_{n_j}(t)\rangle.
 \end{eqnarray}
 Basing on these  definitions, the invariant is diagonalized as follows
 \begin{eqnarray}
  \hat I(t)|\phi_{n_1,n_2}(t)\rangle&=& \nu\left(n_1+n_2+1\right)|\phi_{n_1,n_2}(t)\rangle.
 \end{eqnarray}
Since the Hamiltonian of the system is time-dependent, the Schr\"odinger equation of the system is 
 \begin{equation}\label{x9}
 i\frac{\partial }{\partial t}\psi(x_1,x_2,t)
=\hat H(t)\psi(x_1, x_2,t),\,\,\,\,\,\, \psi(x_1,x_2,t)\in \mathcal{H}
\end{equation}
where the eigenfunction $\psi(x_1,x_2,t)$ is related  to $\phi (x_1,x_2,t)$ by
\begin{equation}\label{x10}
\psi_{n_1,n_2} (x_1,x_2,t)=e^{i\theta_{n_1,n_2}(t)}\phi_{n_1,n_2}(x_1,x_2,t). 
\end{equation}
Inserting this equation  in (\ref{x9}), one determines the phase function $\theta_{n_1,n_2}(t)$
in the form
\begin{equation}\label{w3}
 \theta_{n_1,n_2}(t)=\int_0^t \langle  \phi_{n_1,n_2} (t')| i\frac{\partial}{\partial t'}-\hat H(t')|\phi_{n_1,n_2}(t')\rangle dt'.
\end{equation}
However, as we pointed out in the previous section,  this system possesses a conserved angular-momentum
\begin{eqnarray}\label{w2}
 \hat L_z&=&\hat x_1\hat p_2-\hat x_2\hat p_1\cr
 &=&i(a_2^\dag a_1-a_1^\dag a_2),
\end{eqnarray}
which commutes with the invariant operator and with the Hamiltonian
\begin{eqnarray}
  [\hat L_z,\hat I(t)]=0,\,\,\,\,\,\,\,\,\,[\hat L_z,\hat  H(t)]=0
 \end{eqnarray}
 Although the operator $\hat L_z$  commutes with both $ \hat I(t)$ and $ \hat H(t)$,  the basis $|\phi_{n_1,n_2}(t)\rangle$  cannot diagonalize them simultaneously.
 Therefore, it is convenient to find another basis of Hilbert space that  diagonalizes these   operators.

\section{ Eigensystems of the Hamiltonian operator}\label{sec4}

To recover the available eigenbasis of the invariant operator which can diagonalize simultaneously the invariant operator, the angular momentum
and the Hamiltonian of the system, let us consider  the helicity  Fock algebra generators as follows
 \begin{eqnarray}\label{v1}
  a_{\pm}'&=&\frac{1}{\sqrt{2}}\left(a_1'\pm ia_2'\right),\,\,\, a_{\pm}'^\dag=\frac{1}{\sqrt{2}}\left(a_1'^\dag\mp ia_2'^\dag\right),
 \end{eqnarray}
with 
\begin{eqnarray}
 [a_\pm',a_\pm'^\dag]=\mathbf{I},\,\,\,\,\,\, [a_\pm',a_\mp'^\dag]=0,
\end{eqnarray}
where $a_1', a_2',{a_1'}^\dag, {a_2'}^\dag$ are  the ones in the  previous  equations.
The associated helicity-like basis $|\phi_{n_+,n_-}'(t)\rangle$  are defined as follows
\begin{eqnarray}\label{v2}
 |\phi_{n_+,n_-}'(t)\rangle&=&\frac{1}{\sqrt{n_+!n_-!}}\left(a_+'^\dag\right)^{n_+} \left(a_-'^\dag\right)^{n_-}|\phi_{0,0}'(t)\rangle,\\
 \langle \phi_{n_+,n_-}'(t)|\phi_{m_+,m_-}'(t)\rangle&=&\delta_{n_+,m_+}\delta_{n_-,m_-},
\end{eqnarray}
with $|\phi_{0,0}'(t)\rangle$ is a normalized state annihilated by $ a_\pm' $ as by $a_1',a_2'$.

In order to find the exact expression of the joint eigenfunction of  the invariant operator and the angular momentum, we introduce the polar coordinates 
through the following canonical transformation  $\hat x_1=r\cos\alpha,\,\,\hat x_2=r\sin\alpha,\,
\hat p_1=-i(\cos\alpha \partial_r-\frac{\sin\alpha}{r}\partial_\alpha)$ and
$\hat p_2=-i(\sin\alpha \partial_r+\frac{\cos\alpha}{r}\partial_\alpha)$.
In terms of these coordinates the operators  in equation (51) can be written as
\begin{eqnarray}
 {a'_\pm}^\dag&=&\frac{1}{2}e^{\mp i\alpha}\left[\left(\frac{\nu}{\rho}r-\rho\partial_r\right)\pm i\frac{\rho}{r}\partial_\alpha\right],\\
 a'_\pm&=& \frac{1}{2}e^{\pm i\alpha}\left[\left(\frac{\nu}{\rho}r+\rho\partial_r\right)\mp i\frac{\rho}{r}\partial_\alpha\right].
\end{eqnarray}
 From the relation (\ref{v2}) we construct the eigenfunction for the invariant operator of the system   according to \cite{50}. One finds 
\begin{eqnarray}\label{v}
\phi_{n_+,n_-}(x_1,x_2,t)&=&U^\dag \phi_{n_+,n_-}'(x_1,x_2,t),
\end{eqnarray}
that is
\begin{eqnarray}
\phi_{n_+,n_-}(x_1,x_2,t)&=&(-)^n\frac{(\nu)^{\frac{1+|\ell|}{2}}}{\rho^{1+|\ell|}\sqrt{\pi}}\sqrt{\frac{n!}{\Gamma(n+|\ell|+1)}}r^{|\ell|}e^{\left(imf^{-1}
\frac{\dot{\rho}}{\rho}-\frac{\nu}{\rho^2}\right)\frac{r^2}{2}} \cr&&\times L_n^{|\ell|}\left(\frac{\nu}{\rho^2}r^2\right)e^{i\ell\alpha},
\end{eqnarray}
where $\ell=n_+-n_-$,\,\,\,$n=\min(n_+,n_-)=\frac{1}{2}(n_++n_--|\ell|)$, $\Gamma(u)$ the Gamma function and $L_n^{|\ell|}\left(u\right)$
are the generalised Laguerre polynomials.

To obtain the expectation values of the operators $\hat I(t), \hat L_z, \hat H(t)$ that are  respectively 
$E_{n_\pm},l_{n_\pm},\mathcal{E}_{n_\pm}$, we introduce a new
pair of raising and lowing helicity  operators define as
\begin{eqnarray}
a_\pm&=&U^\dag a_\pm' U= \frac{1}{2\sqrt{\nu}}\left[\left(mf^{-1}\dot{\rho}+i\frac{\nu}{\rho}\right)
\left(\hat x_1\pm i\hat x_2\right)-\rho\left(\hat p_1\pm i\hat p_2\right)\right],\\
a_\pm^\dag&=&U^\dag a_\pm'^\dag U=\frac{1}{2\sqrt{\nu}}\left[\left(mf^{-1}\dot{\rho}-i\frac{\nu}{\rho}\right)
\left(\hat x_1\mp i\hat x_2\right)-\rho\left(\hat p_1\mp i\hat p_2\right)\right].
\end{eqnarray}
In term of these operators we have
\begin{eqnarray}
 \hat I(t)&=& \nu\left(a_+^\dag a_++a_-^\dag a_-+\mathbf{I}\right),\\
 \hat L_z&=&\left(a_-^\dag a_--a_+^\dag a_+\right),\\
 \hat H(t)&=&\frac{1}{2\nu}\left(mf^{-1}\dot{\rho}^2+\frac{f\nu^2}{m\rho^2}+m\omega^2f^{-1}\rho^2\right)\left(a_+^\dag a_+ +a_-^\dag a_-+\mathbf{I}\right)+\cr&&
    \left(-\frac{mf^{-1}\dot{\rho}}{2\nu}+i\frac{\dot{\rho}^2}{\rho}+\frac{f\nu}{2m\rho^2}-\frac{m\omega^2f^{-1}\rho^2}{2\nu}\right)a_-a_++\cr&&
    \left(-\frac{mf^{-1}\dot{\rho}}{2\nu}-i\frac{\dot{\rho}^2}{\rho}+\frac{f\nu}{2m\rho^2}-\frac{m\omega^2f^{-1}\rho^2}{2\nu}\right)a_-^\dag a_+^\dag.
\end{eqnarray}
The expectation values of the above operators read  as
\begin{eqnarray}
 E_{n_\pm}&=&\langle \phi_{n_+,n_-}(t)|\hat I(t)|\phi_{n_+,n_-}(t)\rangle=\nu\left(n_++n_-+1\right), \\
 l_{n_\pm}&=&\langle \phi_{n_+,n_-}(t)|\hat L_z|\phi_{n_+,n_-}(t)\rangle=n_--n_+,\\
 \mathcal{E}_{n_\pm}&=&\langle \phi_{n_+,n_-}(t)|\hat H(t)|\phi_{n_+,n_-}(t)\rangle=
 \frac{1}{2\nu}\left(mf^{-1}\dot{\rho}^2+\frac{f\nu^2}{m\rho^2}+m\omega^2f^{-1}\rho^2\right) \cr&&\times
 \left(n_+ +n_-+1\right),\label{af}
\end{eqnarray}
where the action of $a_\pm$ and $a_\pm^\dag$ on $ |\phi_{n_\pm}(t)\rangle$  finds expression in 
\begin{eqnarray}
  a_\pm^\dag|\phi_{n_\pm,n_\mp}(t)\rangle&=&\sqrt{n_\pm+1}|\phi_{n_\pm+1,n_\mp}(t)\rangle,\\
  a_\pm|\phi_{n_\pm,n_\mp}(t)\rangle&=& \sqrt{n_\pm}|\phi_{n_\pm-1,n_\mp}(t)\rangle,\\
  a_\pm^\dag a_\pm|\phi_{n_\pm,n_\mp}(t)\rangle&=& n_\pm|\phi_{n_\pm,n_\mp}(t)\rangle.
 \end{eqnarray}
 
 To determine the exact solution of the Schr\"odinger equation (\ref{x9}), we have to find the exact
expression of the phase function in equation (\ref{w3}) such that
\begin{eqnarray}\label{t7}
 \frac{d}{dt}\theta_{n_1,n_2}(t)&=&\langle \phi_{n_+,n_-}(t)|i\frac{\partial }{\partial t}-\hat H(t)|\phi_{n_+,n_-}(t)\rangle\cr
                      &=& \langle \phi_{n_+,n_-}(t)|i\frac{\partial }{\partial t}|\phi_{n_+,n_-}(t)\rangle-
                      \langle \phi_{n_+,n_-}(t)|\hat H(t)|\phi_{n_+,n_-}(t)\rangle.
                      \end{eqnarray}
Let us evaluate the following expression
\begin{eqnarray}\label{w4}
   \langle \phi_{n_+,n_-}(t)|\frac{\partial}{\partial t}|\phi_{n_+,n_-}(t)\rangle&=&\frac{1}{\sqrt{n_+!n_-!}}
    \langle \phi_{n_+,n_-}(t)|\frac{\partial}{\partial t}\left[\left(a_+^\dag\right)^{n_+} \left(a_-^\dag\right)^{n_-}|\phi_{0,0}(t)\rangle\right]\cr
    &=&\langle \phi_{0,0}(t)|\frac{\partial}{\partial t}|\phi_{0,0}(t)\rangle+\frac{1}{\sqrt{n_+!n_-!}}\cr&&\times
     \langle \phi_{n_+,n_-}(t)|\frac{\partial}{\partial t}\left[\left(a_+^\dag\right)^{n_+} \left(a_-^\dag\right)^{n_-}\right]|\phi_{0,0}(t)\rangle.
  \end{eqnarray}
  We have
\begin{eqnarray}\label{w5}
 \langle \phi_{0,0}(t)|\frac{\partial}{\partial t}|\phi_{0,0}(t)\rangle=\frac{imf^{-1}}{2\nu}(\ddot{\rho}\rho+\dot{\rho}\rho-\dot{\rho}^2),
\end{eqnarray} 
and 
\begin{eqnarray}\label{w6}
 \frac{1}{\sqrt{n_+!n_-!}}
\langle \phi_{n_+,n_-}(t)|\frac{\partial}{\partial t}\left[\left(a_+^\dag\right)^{n_+} \left(a_-^\dag\right)^{n_-}\right]|\phi_{0,0}(t)\rangle&=&
\frac{imf^{-1}}{2\nu}\left(\ddot{\rho}\rho+\eta\dot{\rho}\rho-\dot{\rho}^2\right)\cr&&\times(n_++n_-),
\end{eqnarray}
where the expressions of $\frac{\partial a_+^\dag}{\partial t}$ and $\frac{\partial a_-^\dag}{\partial t}$ in terms of $a_\pm$ and $a_\pm^\dag$ are
\begin{eqnarray}
 \frac{\partial a_+^\dag}{\partial t}&=& \frac{1}{2\sqrt{\nu}}\left[\left(m f^{-1}\eta\dot{\rho}+m f^{-1}\ddot{\rho}+i\nu\frac{\dot{\rho}}{\rho^2}
 \right)(\hat x_1-i\hat x_2)-\dot{\rho}(\hat p_1-i\hat p_2)\right]\cr
 &=&\frac{imf^{-1}}{2\nu}\left(\ddot{\rho}\rho+\eta\dot{\rho}\rho-\dot{\rho}^2\right)a_+^\dag+\left[\frac{\dot{\rho}}{\rho}-\frac{imf^{-1}}{2\nu}
 \left(\ddot{\rho}\rho+\eta\rho-\dot{\rho}^2\right)\right]a_-,\\
 \frac{\partial a_-^\dag}{\partial t}&=& \frac{1}{2\sqrt{\nu}}\left[\left(m f^{-1}\eta\dot{\rho}+m f^{-1}\ddot{\rho}+i\nu\frac{\dot{\rho}}{\rho^2}
 \right)(\hat x_1+i\hat x_2)-\dot{\rho}(\hat p_1+i\hat p_2)\right]\cr
   &=&\frac{imf^{-1}}{2\nu}\left(\ddot{\rho}\rho+\eta\dot{\rho}\rho-\dot{\rho}^2\right)a_-^\dag+\left[\frac{\dot{\rho}}{\rho}-\frac{imf^{-1}}{2\nu}
 \left(\ddot{\rho}\rho+\eta\rho-\dot{\rho}^2\right)\right]a_+.
\end{eqnarray}
We then find
\begin{eqnarray}\label{w7}
  \langle \phi_{n_+,n_-}(t)|\frac{\partial}{\partial t}|\phi_{n_+,n_-}(t)\rangle&=& 
  \frac{imf^{-1}}{2\nu}\left(\ddot{\rho}\rho+\eta\dot{\rho}\rho-\dot{\rho}^2\right)(n_++n_-+1)\cr
  &=&\frac{imf^{-1}}{2\nu}\left(\frac{\nu^2f^2}{m^2\rho^2}-\omega^2\rho^2-\dot{\rho}^2\right)(n_++n_-+1).
\end{eqnarray}
Finally, taking into account (\ref{af}) and (\ref{w7}), we find that the phase function in (\ref{t7})  is given
by
\begin{equation}
 \theta_{n_+,n_-}(t)=-\frac{\nu}{2m}(n_++n_-+1)\int_0^t\frac{f(t')}{\rho^2(t')}dt'.
\end{equation}
Our  result for $\theta_{n_+,n_-}(t)$  confirms the $2$D case result of \cite{26}, slightly differs  from the one calculated in \cite{14} and 
largerly differs from our previous  result \cite{51}. In fact,  in the presence of  an external electromagnetic field $[\vec B(\vec x,t),
\vec E(\vec x,t)]$, we obtain the phase function of \cite{26} due to the contribution of 
the magnetic field  $\vec B(\vec x,t)$ which  induces a minimal coupling   in the Hamiltonian $\hat H(\vec x,\vec p,t)$.
Indeed, in addition to the magnetic field contribution, with an appropriate canonical and gauge  transformations on the 
   electric field $\vec E(\vec x,t)$, this phase function is extended to the one obtained in  \cite{51}.
   
The solution of the Schr\"odinger equation is given by
\begin{eqnarray}\label{vv}
\psi_{n, \ell}(x_1,x_2,t)&=& (-)^n\frac{(\nu)^{\frac{1+|\ell|}{2}}}{\rho^{1+|\ell|}\sqrt{\pi}}
\sqrt{\frac{n!}{\Gamma(n+|\ell|+1)}}r^{|\ell|}e^{\left(imf^{-1}
\frac{\dot{\rho}}{\rho}-\frac{\nu}{\rho^2}\right)\frac{r^2}{2}}\cr&& \times L_n^{|\ell|}\left(\frac{\nu}{\rho^2}r^2\right)e^{i\ell\alpha}e^{i\theta_{n,\ell}(t)}.
\end{eqnarray}

However, one can deduce from the Lagrangian (\ref{eq1}) the usual kinetic momentum $p_{k_j}$ such as
\begin{eqnarray}
 p_{k_j}=\frac{\partial L}{\partial \dot{x}_j}=f(t)p_j,\,\,\,\,\,\,\,j=1,2,
\end{eqnarray}
 where $p_j$ the canonical momentum and $ p_{k_j}=m\dot{x}_j$. The mechanical energy of the system in term of the Hamiltonian (\ref{x1}) reads as
 \begin{eqnarray}\label{E_m}
  E_m&=&\frac{m}{2}\dot{x}_j^2+\frac{m\omega^2(t)}{2}x_j^2\cr
  &=&f(t) H(t).
 \end{eqnarray}
As pointed out in the literature by several authors \cite{52,53,54,55,56},
the quantization of this dissipative system  for particular value of the function 
$f(t)=e^{-\gamma t}$ through a non-inertial canonical transformation, is unsatisfactory with the laws of quantum theory such  that
 the zero-point of the expectation values of the energy  instead of going to the quantum  ground energy 
 and the  violation of  the Heisenberg uncertainty relations when one tends the time to infinity
($t\rightarrow\infty$). Therefore, the expectation value of the mechanical energy (\ref{E_m}) is given by
\begin{eqnarray}
 \langle\psi_{n,\ell}| E_m|\psi_{n,\ell}\rangle&=&\frac{1}{2\nu}\left(m\dot{\rho}^2+
 \frac{f^2\nu^2}{m\rho^2}+m\omega^2\rho^2\right)\left(2n+|\ell|+1\right)
\end{eqnarray}
 and 
\begin{eqnarray}
 \lim_{t\rightarrow\infty}\langle\psi_{n,\ell}| E_m|\psi_{n,\ell}\rangle\neq0,\,\,\,
 \forall f\in \mathbb R
\end{eqnarray}
One infers that the problem  of the zero-point  energy caused  by the  use of the non-inertial  canonical transformation 
is raised up by this method of Lewis-Riesenfeld.  In the next section, let us check  the validity 
of the generalized version of  the  Heisenberg's uncertainty relations.

\section{Heisenberg's uncertainty relations}\label{sec5}

To prove the validity of the   generalized uncertainty relations (\ref{He}) with $\hbar=1$, we  start
with the determination of the standard expectation values of the  operators $\hat x_1,\hat x_2,\hat p_1,\hat p_2$ and $\hat p_{k_j}$
\begin{eqnarray}
 \langle \psi_{n,\ell}|\hat x_1|\psi_{n,\ell}\rangle&=& \langle \psi_{n,\ell}|\hat x_2|\psi_{n,\ell}\rangle=0,\\
  \langle \psi_{n,\ell}|\hat p_1|\psi_{n,\ell}\rangle&=& \langle \psi_{n,\ell}|\hat p_2|\psi_{n,\ell}\rangle=0,\\
   \langle \psi_{n,\ell}|\hat x_1^2|\psi_{n,\ell}\rangle&=&\langle \psi_{n,\ell}|\hat x_2^2|\psi_{n,\ell}\rangle= \frac{\rho^2}{2\nu}\left(2n+|\ell|+1\right),\\
\langle \psi_{n,\ell}|\hat p_1^2|\psi_{n,\ell}\rangle&=&\langle \psi_{n,\ell}|\hat p_2^2|\psi_{n,\ell}\rangle=
\left(2n+|\ell|+1\right) \left(\frac{m^2f^{-2}\dot{\rho}^2}{2\nu}+\frac{\nu}{2\rho^2}\right),\\
    \langle \psi_{n,\ell}|[\hat x_1,\hat p_1]|\psi_{n,\ell}\rangle&=& \langle n,\ell|[\hat x_2,\hat p_2]|n,\ell\rangle= i,\\
    \langle \psi_{n,\ell}|[\hat x_1,\hat p_{k_1}]|\psi_{n,\ell}\rangle&=& \langle n,\ell|[\hat x_2,\hat p_{k_2}]|n,\ell\rangle= if(t).
  \end{eqnarray}
The dispersions  of operators are computed to
\begin{eqnarray}
 \Delta x_1=\Delta x_2&=&\sqrt{\frac{\rho^2}{2\nu}\left(2n+|\ell|+1\right)},\\
 \Delta p_1=\Delta p_2&=&\sqrt{\frac{1}{2}\left(2n+|\ell|+1\right) \left(\frac{m^2f^{-2}\dot{\rho}^2}{\nu}+\frac{\nu}{\rho^2}\right)},\\
 \Delta p_{k_1}=\Delta p_{k_2}&=& f(t)\sqrt{\frac{1}{2}\left(2n+|\ell|+1\right) \left(\frac{m^2f^{-2}\dot{\rho}^2}{\nu}+\frac{\nu}{\rho^2}\right)}.
\end{eqnarray}
The Heisenberg uncertainty relations   can be inferred 
 \begin{eqnarray}
  \Delta x_1 \Delta p_1&=&\Delta x_2 \Delta p_2=\frac{1}{2}\left(2n+|\ell|+1\right)\sqrt{1+\frac{m^2f^{-2}\dot{\rho}^2\rho^2}{\nu^2}}\geq \frac{1}{2},\\
  \Delta x_1 \Delta p_{k_1}&=&\Delta x_2 \Delta p_{k_2}=\frac{f(t)}{2}\left(2n+|\ell|+1\right)
  \sqrt{1+\frac{m^2f^{-2}\dot{\rho}^2\rho^2}{\nu^2}}\geq \frac{f(t)}{2},\label{p1}\\
 \Delta x_1\Delta x_2&=& \frac{\rho^2}{2\nu}\left(2n+|\ell|+1\right)\geq 0,\\
  \Delta p_1\Delta p_2&=& \left(2n+|\ell|+1\right) \left(\frac{m^2f^{-2}\dot{\rho}^2}{2\nu}+\frac{\nu}{2\rho^2}\right)\geq 0,\\
  \Delta p_{k_1}\Delta p_{k_2}&=& f^2(t)\left(2n+|\ell|+1\right) \left(\frac{m^2f^{-2}\dot{\rho}^2}{2\nu}+\frac{\nu}{2\rho^2}\right)\geq 0.
 \end{eqnarray}
 These results are related to similar discussions in \cite{10}. In the present case 
 the uncertainty relations are satisfied except  for the relation in equation (\ref{p1}). In fact
 this uncertainty relation  may   tend to zero if  $\lim_{t\rightarrow \infty}f(t)\rightarrow 0$ (for instance $f(t)=e^{-\gamma t}).$
 This result seems to violate the Heisenberg uncertainty relations, 
but  as observed authors in \cite{52,53,54,55,56}, this result cannot disagree with the quantum mechanics theory,
because the uncertainty relations hold only for the conjugate canonical operators $\hat x_j$ and $\hat p_j$. 
Accordingly, the Lewis-Riesenfeld approach removes all 
the  major objections related to this model.

As we can also remark, with this approach, the determination of the spectrum allowed the introduction of 
the nonstationary discrete eigenbasis. Thus, to convert this spectrum into nonstationary 
continuous spectrum, it is useful to introduce a continuous   basis in which the diagonalization is possible. In this sense, the  coherent 
states are the best candidates to achieve this purpose. In the literature, various  coherent states \cite{57,58,59} are contructed for different Lie algebra.
To construct  the appropriate coherent states for this system whose
 eigenfunction is  expressed in terms of the generalized Laguerre functions as in \cite{61,62,63,64,65,66}, we factorise this eigenfunction 
 to find the hidden symmetry of the
system through the establishment of an appropriate Lie algebra.

\section{The hidden dynamical Lie  algebra}\label{sec6}

We construct in this section  the raising and lowering operators from the Hamiltonian's eigenfunction  which generate the hidden Lie algebra. 
Since  the eigenfunctions of the invariant operator and  the Hamiltonian  are expressed in terms of the generalized Laguerre
functions $L_n^{\ell}(u)$ with $\ell>0$.
It is important to review some useful properties  related to this  special function that will be used to generate the symmetry operators.
 Thus, the generalized Laguerre polynomials $L_n^{\ell}(u)$  are defined  as \cite{67}
\begin{eqnarray}
 L_n^{\ell}(u)&=&\frac{1}{n!}e^u u^{-\ell}\frac{d^n}{du^n}(e^{-u}u^{n+\ell}).
  \end{eqnarray}
  For $\ell=0,\,\,\,L_n^0(u)=L_n(u)$  and \,\,\,for $n=0,\,\,\,L_0^{\ell}(u)=1$. The generating functions corresponding to associated
  Laguerre polynomials are
  \begin{eqnarray}
   \frac{e^{\frac{uz}{z-1}}}{(1-z)^{1+\ell}}&=&\sum_{n=0}^\infty L_n^{\ell}(u)z^n,\,\,\,\,\,\,|z|<1,\label{g}\\
 J_{\ell}\left(2\sqrt{uz}\right)e^z(uz)^{-\frac{\ell}{2}}&=&\sum_{n=0}^\infty\frac{z^n}{\Gamma(n+\ell+1)}L_n^{\ell}(u)\label{j},
\end{eqnarray}
where the $ J_\kappa(x)$ is the ordinary Bessel function of $\kappa$-order.

The orthogonality relation is 
\begin{eqnarray}
 \int_0^{\infty}du e^{-u}u^{\ell}L_n^{\ell}(u)L_m^{\ell}(u)= \frac{\Gamma(\ell+n+1)}{n!} \delta_{nm}.                                                     
\end{eqnarray}
 The generalised Laguerre polynomials satisfy the following differential equation
 \begin{equation}
  \left[u\frac{d^2}{du^2}+(\ell-u+1)\frac{d}{du}+n\right]L_n^{\ell}(u)=0,
 \end{equation}
and the recurrence relations
\begin{eqnarray}
 (n+1)L_{n+1}^{\ell}(u)-\left(2n+\ell+1-u\right)L_n^{\ell}(u)+\left(n+\ell\right)L_{n-1}^{\ell}(u)&=&0,\label{r1}\\
 u\frac{d}{du}L_n^{\ell}(u)-nL_n^{\ell}(u)+(n+\ell)L_{n-1}^{\ell}(u)&=&0.\label{r2}
\end{eqnarray}
With respect to the  equations, we rewrite the  eigenfunction of the invariant operator in equation (\ref{ajou}) in the form
\begin{equation}
 \phi_n^{\ell}(u)=N(\rho,\alpha)\sqrt{\frac{n!}{\Gamma(n+\ell+1)}}u^{\frac{\ell}{2}}e^{-\frac{\varpi}{2}u}L_n^{\ell}(u),
\end{equation}
where $u=\frac{\nu}{\rho^2}r^2$, $N(\rho,\alpha)=(-)^n\sqrt{\frac{\nu}{\pi\rho^2}}e^{i\ell\alpha}$,\,\,\,\,
$\varpi=1-imf^{-1}\frac{\rho\dot{\rho}}{\nu}$ and $\Gamma(n)=(n-1)!$.\\
Basing on the recurrence relations (\ref{r1}) and (\ref{r2}), we obtain the following equations
\begin{eqnarray}
 \left(-u\frac{d}{du}+\frac{\ell}{2}+n-\frac{\varpi}{2}u\right)\phi_n^{\ell}(u)&=&\sqrt{n(n+\ell)}\phi_{n-1}^{\ell}(u),\\
 \left(u\frac{d}{du}+\frac{\ell}{2}+n-\frac{\tilde \varpi}{2}u+1\right)\phi_n^{\ell}(u)&=&\sqrt{(n+1)(n+\ell+1)}\phi_{n+1}^{\ell}(u),
\end{eqnarray}
where $\tilde \varpi=2-\varpi$.
For the sake of simplicity we define the raising operator $K_+$ and the lowering operator $K_-$ acting on the wave function $ \phi_n^{\ell}(u)$ as
\begin{eqnarray}
K_-=\left(-u\frac{d}{du}+\frac{\ell}{2}+n-\frac{\varpi}{2}u\right),\\
K_+=\left(u\frac{d}{du}+\frac{\ell}{2}+n-\frac{\tilde \varpi}{2}u+1\right),
\end{eqnarray}
 and  hence obtain
 \begin{eqnarray}
  K_-\phi_n^{\ell}(u)&=&\sqrt{n(n+\ell)}\phi_{n-1}^{\ell}(u),\\
  K_+\phi_n^{\ell}(u)&=&\sqrt{(n+1)(n+\ell+1)}\phi_{n+1}^{\ell}(u).
 \end{eqnarray}
By multiplying  both side of the latter  equations  by the factor $e^{i\theta_{n,\ell}(t)}$  we obtain
\begin{eqnarray}
 K_-\psi_n^{\ell}(u)&=& \sqrt{n(n+\ell)}\psi_{n-1}^{\ell}(u),\label{s}\\
  K_+\psi_n^{\ell}(u)&=&\sqrt{(n+1)(n+\ell+1)}\psi_{n+1}^{\ell}(u).
\end{eqnarray}
 By successively applying $K_+$ on the ground state $\psi_0^{\ell} (u)$, we generate the eigenfunction  $\psi_n^{\ell} (u)$ of the system 
 as follows
 \begin{eqnarray}
  \psi_n^{\ell}(u)&=&\sqrt{\frac{\Gamma(1+\ell)}{n!\Gamma(n+\ell+1)}}(K_+)^n\psi_0^{\ell} (u),\\
 \end{eqnarray}
 where,
 \begin{eqnarray}
 \psi_0^{\ell}(u)&=&\frac{N(\rho,\alpha)}{\sqrt{\Gamma(\ell+1)}}u^{\frac{\ell}{2}}e^{-\frac{\varpi}{2}u}e^{i\theta_{n,\ell}(t)},\\
 K_- \psi_0^{\ell}(u)&=&0.
 \end{eqnarray}
 One can also observe that the following relations are satisfied
\begin{eqnarray}
 K_+K_-\psi_n^{\ell} (u)&=&n(n+\ell)\psi_n^{\ell} (u),\\
 K_+K_-\psi_n^{\ell} (u)&=&(n+1)(n+\ell+1)\psi_n^{\ell} (u).
\end{eqnarray}
Now, to establish the dynamical Lie algebra associated with the ladder operators $K_\pm$, we calculate the
commutator
\begin{equation}
 [K_-,K_+]\psi_n^{\ell}(u)=(2n+\ell+1)\psi_n^{\ell}(u).
\end{equation}
As a consequence, we can introduce the operator $K_0$ defined to satisfy
\begin{eqnarray}
 K_0\psi_n^{\ell}(u)=\frac{1}{2}(2n+\ell+1)\psi_n^{\ell}(u).
\end{eqnarray}
The operators $K_\pm$ and $K_0$  satisfy the following  commutation relations
\begin{eqnarray}
 [K_-,K_+]=2K_0,\,\,[K_0,K_\pm]=\pm K_\pm,
\end{eqnarray}
which can be recognized as commutation relation of the generators of a non-compact Lie
algebra $su(1,1)$. The corresponding Casimir operator for any irreducible representation is the identity times a number
\begin{eqnarray}
 K^2=K_0^2-\frac{1}{2}(K_+K_-+K_-K_+)=\frac{1}{4}(\ell+1)(\ell-1).
\end{eqnarray}
It satisfies
\begin{equation}
 [K^2,K_\pm]=0=[K^2,K_0].
\end{equation}
If we make the following   connection between  the physical quantum numbers $(n,\ell)$ and the  ordinary   $SU(1,1)$   group numbers $(n,k)$ such as               
\begin{eqnarray}
 \ell=2k-1,   
\end{eqnarray}
  then we recover the ordinary discrete representations of the $su(1,1)$ Lie algebra 
  \begin{eqnarray}
   K^2\psi_n^{k}(u)&=&k(k-1)\psi_n^{k}(u),\\
   K_-\psi_n^{k}(u)&=& \sqrt{n(n+2k-1)}\psi_{n-1}^{k}(u),\label{s}\\
  K_+\psi_n^{k}(u)&=&\sqrt{(n+1)(n+2k)}\psi_{n+1}^{k}(u),\\   
  K_0\psi_n^{k}(u)&=& (n+k)\psi_n^{k}(u).
  \end{eqnarray}
Thus, in what   follows   we use the Bargmann index $\ell$ instead of the ordinary  index $k$ in  the  representation of $su(1, 1)$ algebra. 
Now, with the properties of the generators $K_\pm$ and $K_0$ of the $su(1,1)$ algebra, we are in the position to construct the corresponding
coherent states  to this system.

\section{SU(1,1) coherent states}\label{sec7}

We investigate in this section the $SU(1,1)$ coherent states by adopting  Barut-Girardello \cite{30} and Perelomov \cite{31} approaches. We examin for each
approach the resolution of identity and overlapping properties.

\subsection{Barut-Girardello coherent states}

\subsubsection{Construction}
Following the Barut and Girardello approach \cite{30},  $ SU(1,1)$ coherent states  are defined  to  be 
the eigenstates of the lowering generator $K_-$
\begin{equation}\label{Bar1}
 K_-|\psi_z^\ell\rangle=z |\psi_z^\ell\rangle,
\end{equation}
where $z$ is an arbitrary complex number. Based on the completeness of the wavefunction such that
$\sum_{n=0}^\infty|\psi_n^{\ell}\rangle\langle \psi_n^{\ell}|=\mathbf{I}$, on
can represent the coherent states $|\psi_z^\ell\rangle$ as follows
\begin{eqnarray}\label{Bar2}
 |\psi_z^\ell\rangle&=& \sum_{n=0}^\infty \langle \psi_n^{\ell}|\psi_z^\ell\rangle|\psi_n^{\ell}\rangle.
\end{eqnarray}
Acting the operator $K_-$ on the equation (\ref{Bar2}) and then, using the equations (\ref{Bar1}) and (\ref{s})
we have the following result
\begin{eqnarray}
 \langle \psi_n^{\ell}|\psi_z^\ell\rangle=\frac{z}{\sqrt{n(n+\ell)}}\langle \psi_{n-1}^{\ell}|\psi_z^\ell\rangle.
\end{eqnarray}
After the recurrence procedure,  the formal equation becomes
\begin{equation}
 \langle \psi_n^{\ell}|\psi_z^\ell\rangle=z^n\sqrt{\frac{\Gamma(1+\ell)}{n!\Gamma(n+\ell+1)}}\langle \psi_0^{\ell}|\psi_z^\ell\rangle.
\end{equation}
Referring to \cite{67}, the Gamma function is linked to the modified Bessel function $I_\mu(x)$ of order $\mu$ through the relation
\begin{equation}
 \sum_{n=0}^\infty\frac{x^{2n}}{n!\Gamma(n+\mu+1)}=\frac{I_\mu(2x)}{x^\mu}.
\end{equation}
Therefrom, by setting $x=z$  and  $\mu=\ell$,  we deduce the Barut-Girardello coherent states as fallows
\begin{eqnarray}
 |\psi_z^\ell\rangle&=&
 \sqrt{\frac{|z|^{\ell}}{I_{\ell}(2|z|)}}\sum_{n=0}^\infty\frac{z^n}{\sqrt{n!\Gamma(n+\ell+1)}}|\psi_n^{\ell}\rangle,\label{cbar}\\
 \psi_z^{\ell}(u)&=&\frac{|z|^{\frac{\ell}{2}} N(\rho,\alpha)}{\sqrt{I_{\ell}(2|z|)}}
\sum_{n=0}^\infty \frac{z^n}{\Gamma(n+\ell+1)}u^{\frac{\ell}{2}}e^{-\frac{\varpi}{2}u}L_n^{\ell}(u)e^{i\theta_{n,\ell}(t)}.
\end{eqnarray}

However, in term of the generating function  (\ref{j}), the Barut-Girardello coherent  states can be written as follows
\begin{eqnarray}
  \psi_z^{\ell}(u)&=&\left(\frac{z}{|z|}\right)^{-\frac{\ell}{2}}\frac{ N(\rho,\alpha) e^{z-\frac{\varpi}{2}u} }{\sqrt{I_{\ell}(2|z|)}}
 J_{\ell}\left(2\sqrt{uz}\right)e^{i\theta_{n,\ell}(t)}.
\end{eqnarray}

\subsubsection{Properties}

It is well-known  that the states (\ref{cbar}) are normalized but not orthogonal and satisfy the resolution of identity. Thus, we can see that the 
scalar product of two coherent states does not vanish
\begin{equation}
 \langle\psi_{z_1}^\ell|\psi_{z_2}^\ell\rangle=\frac{I_{\ell}(2\sqrt{ z_1^*z_2})}{\sqrt{I_{\ell}(2|z_1|)|I_{\ell}(2|z_2|)}}.
\end{equation}
The overcompleteness relation reads  as follows
\begin{equation}
 \int d\mu(z,\ell)|\psi_z^\ell\rangle\langle \psi_z^\ell|=\sum_{n=0}^\infty|\psi_n^{\ell}\rangle\langle \psi_n^{\ell}|=\mathbf{I},
\end{equation}
with the measure
\begin{equation}
 d\mu(z,\ell)=\frac{2}{\pi}K_{\ell}(2|z|)I_{\ell}(2|z|)d^2z,
\end{equation}
where $d^2z=d(Re z)d(Im z)$ and $K_\upsilon(x)$ is the $\upsilon$-order modified Bessel function of the second kind.\\
For arbitrary  state $|\Phi\rangle=\sum_{n=0}^\infty c_n|\psi_n^{\ell}\rangle$ in the Hilbert space, one can construct  the analytic function $f(z)$
such as
\begin{eqnarray}
 f(z)=\sqrt{\frac{I_{\ell}(2|z|)}{|z|^{\ell}}}\langle \psi_z^\ell|\Phi\rangle=\sum_{m=0}^\infty\frac{c_m}{\sqrt{m!\Gamma(m+\ell+1)}}z^m.
\end{eqnarray}
On the Barut-Girardello coherent states (\ref{cbar}) one  can explicitly express the state $|\Phi\rangle$ as follows
\begin{equation}
 |\Phi\rangle=\int d\mu(z,\ell)\frac{({z^*})^{\frac{\ell}{2}}}{\sqrt{I_{\ell}(2|z|)}}f(z)|\psi_z^\ell\rangle,
\end{equation}
 and  we have
 \begin{equation}
  ||\Phi||^2=\int d\mu(z,\ell)\frac{|z|^{\ell}}{I_{\ell}(2|z|)}|f(z)|^2<\infty.
 \end{equation}
 
\subsection{Perelomov coherent states}

\subsubsection{Construction}

 In analogy   to canonical   coherent states   construction, Perelomov $SU(1,1)$
 coherent states $|\psi_\eta^\ell\rangle$ are obtained by acting  the displacement operator  $S(\xi)$ on the ground state 
$|\psi_0^{\ell}\rangle$ \cite{31}
\begin{eqnarray}
 |\psi_\eta^\ell\rangle&=& S(\xi)|\psi_0^{\ell}\rangle,\cr
                &=& \exp\left(\xi K_+-\xi^*K_-\right)|\psi_0^{\ell}\rangle,
\end{eqnarray}
where $\xi\in\mathbb C,$  such as $\xi= -\frac{\theta}{2}e^{-i\varphi}$, with $-\infty<\theta<+\infty$  and $0\leq\varphi\leq 2\pi$.

Using Baker-Campbell-Haussdorf relation, we explicit the displacement operator as follows \cite{68}
\begin{equation}\label{d}
 S(\xi)= \exp(\eta K_+)\exp(\zeta K_0)\exp(-\eta^* K_-),
\end{equation}
where $\eta=-\tanh(\frac{\theta}{2})e^{-i\varphi}$ and $\zeta=-2\ln\cosh|\xi|=\ln(1-|\eta|^2)$. By using this normal 
form of the  displacement operator  (\ref{d}), the standard Perelomov $SU(1,1)$ coherent states are  found to be
\begin{eqnarray}
 |\psi_\eta^\ell\rangle&=&(1-|\eta|^2)^{\ell+1}\sum_{n=0}^\infty\sqrt{\frac{\Gamma(n+\ell+1)}{n!\Gamma(\ell+1)}}\eta^n|\psi_n^{\ell}\rangle,\label{P}\\
 \psi_\eta^{\ell}(u)&=&N(\rho,\alpha)\frac{(1-|\eta|^2)^{\ell+1}}{\sqrt{\Gamma(\ell+1)}}u^{\frac{\ell}{2}}e^{-\frac{\varpi}{2}u}
 \sum_{n=0}^\infty \eta^n L_n^{\ell}(u) e^{i\theta_{n,\ell}(t)}.
 \end{eqnarray}

In term of the generating function  (\ref{g}), the Perelomov coherent  states can be written as follows
\begin{equation}
 \psi_\eta^{\ell}(u)=N(\rho,\alpha)\frac{(1-|\eta|^2)^{\ell+1}}{\sqrt{\Gamma(\ell+1)}}u^{\frac{\ell}{2}}e^{-\frac{\varpi}{2}u}
 \frac{e^{\frac{u\eta}{\eta-1}}}{(1-\eta)^{1+\ell}}e^{i\theta_{n,\ell}(t)}.
\end{equation}

\subsubsection{Properties}

The Perelomov $SU(1,1)$ coherent states as the Barut-Girardello coherent states are normalized states but not orthogonal 
\begin{equation}
 \langle \psi_{\eta_1}^\ell|\psi_{\eta_2}^\ell\rangle=\left[(1-|\eta_1|^2)(1-|\eta_2|^2)\right]^{\frac{\ell+1}{2}}(1-\eta_1\eta_2^*)^{-\ell-1},
\end{equation}
and satisfy the completeness relation
\begin{eqnarray}
 \int |\psi_\eta^\ell\rangle\langle \psi_\eta^\ell|d\mu(\eta,\ell)=\sum_{n=0}^\infty|\psi_n^{\ell}\rangle\langle \psi_n^{\ell}|=\mathbf{I}
\end{eqnarray}
where the measure $d\mu(\eta,\ell)=\frac{\ell}{\pi}\frac{d^2\eta}{(1-|\eta|^2)^2}$.\\
As we noted for the Barut-Girardello coherent states,  for any $|\Psi\rangle=\sum_{n=0}^\infty c_n|\psi_n^{\ell}\rangle$
in the Hilbert space, one can construct  an analytic function
\begin{eqnarray}
 f(\eta)=(1-|\eta|^2)^{-\ell-1}\langle \psi_\eta^\ell|\Psi\rangle=\sum_{n=0}^\infty c_n
 \sqrt{\frac{\Gamma(n+\ell+1)}{n!\Gamma(\ell+1)}}(\eta^*)^n.
\end{eqnarray}
The expansion of $|\Psi\rangle$ on the basis of coherent states (\ref{P}) can be written as 
\begin{eqnarray}
 |\Psi\rangle &=&\int d\mu(\eta,\ell)(1-|\eta|^2)^{\frac{\ell+1}{2}}f(\eta)|\psi_\eta^\ell\rangle,\\
 ||\Psi||^2&=&\int d\mu(\eta,\ell)(1-|\eta|^2)^{\ell+1}|f(\eta)|^2<\infty.
\end{eqnarray}

\section{Conclusion}\label{sec8}

In this paper we have investigated the system of a nonrelativistic particle   of mass $m$ with  time-dependent  harmonic frequency $\omega(t)$
in  rotational symmetric  in the plane  under the  influence of  a time-dependent friction force.  At  the classical level we 
solved  the   equations of motion which describe  three particulary  physical systems. At the quantum level,   we used the
Lewis-Riesenfeld's method to construct the spectra of  the invariant operator $\hat I(t)$ and the Hamiltonian $\hat H(t)$ on the 
helicity-like basis $|\phi_{n_\pm}(t)\rangle$. The configuration space wave functions of both operators  are  expressed in terms of the
generalized Laguerre polynomials. 
This system previously introduced in the one dimensional case  \cite{26,27} as the generalization of the Kanai Hamiltonian \cite{46}
has been criticized for violating certain laws of quantum theory.  Nevertheless, as many  approaches of solution have been 
given to raise these controversies
\cite{52,53,54,55,56}, we  used the invariant method of Lewis-Riesenfeld to confirm the preservation of those laws by 
investigating the validity  of the 
Heisenberg uncertainty relations and the expectation values of mechanical energy. 
  
This model generalizes not only the $1$D damped systems   studied in the literature \cite{26,27} but also improves 
the technique of quantization of those model achieved in the framework of Lewis-Riesenfeld method \cite{23,24,25,26,27}. 
By analogy with the work of Pedrosa \cite{27} who constructed the canonical coherent states for the $1$D case of this system, 
we constructed the system of  $SU(1,1)$ coherent states based on the eigenfunction of the Hamiltonian. 
For these states the  resolution of identity and some properties are examined. Referring  to the original paper of Lewis-Riesenfeld \cite{1}, 
it would be also good to determine the transition amplitude connecting any initial state in the remote past to any final state
in the remote future in the case of a constant frequency, we hope to report these aspects elsewhere.
 
\section*{Appendix: Expressions of the phase space operators in terms of the helicity Fock algebra generators}

In this appendix we explicitly develop some intermediary calculations  which allowed us to determine the expressions of operators $\hat I (t)$, $\hat L_z$, 
$\hat H(t)$ of section \ref{sec4} and the Heisenberg uncertainty relations of section \ref{sec5}
 \begin{eqnarray}
 a_j&=&\hat U^\dag a'_j \hat U=\frac{1}{\sqrt{2\nu}}\left(mf^{-1}\dot{\rho}\hat x_j-\rho \hat  p_j+i\frac{\nu}{\rho}\hat x_j\right)\label{xx7},\\
 a_j^\dag&=& \hat U^\dag {a'}_j^\dag \hat U=\frac{1}{\sqrt{2\nu}}\left(mf^{-1}\dot{\rho}\hat x_j-\rho \hat p_j-i\frac{\nu}{\rho}\hat x_j\right) \label{xx8}.
\end{eqnarray}
 with $j=1,2$. Conversely,
\begin{eqnarray}
\hat  x_j=\frac{i\rho}{\sqrt{2\nu}}\left(a_j^\dag-a_j\right), \,\,\, \hat p_j=\frac{imf^{-1}\dot{\rho}}{\sqrt{2\nu}}\left(a_j^\dag-a_j\right)
 -\frac{\sqrt{2\nu}}{2\rho}\left(a_j^\dag+a_j\right).
\end{eqnarray}
The helicity  Fock algebra generators in terms of generators $a_j$ and $a_j^\dag$  are given as follows
 \begin{eqnarray}\label{vv1}
  a_{\pm}&=&\frac{1}{\sqrt{2}}\left(a_1\pm ia_2\right),\,\,\, a_{\pm}^\dag=\frac{1}{\sqrt{2}}\left(a_1^\dag\mp ia_2^\dag\right).
 \end{eqnarray}
The inverse relations are,
\begin{eqnarray}
  a_1&=& \frac{1}{\sqrt{2}}\left(a_++a_-\right) ,\,\,\, a_1^\dag=\frac{1}{\sqrt{2}}\left(a_+^\dag+ a_-^\dag\right),\cr
  a_2&=& -\frac{i}{\sqrt{2}}\left(a_+-a_-\right) ,\,\,\,  a_2^\dag=\frac{i}{\sqrt{2}}\left(a_+^\dag- a_-^\dag\right).
\end{eqnarray}
In terms of helicity generators, the phase space operators read
\begin{eqnarray}
 \hat x_1&=&-\frac{i\rho}{2\sqrt{\nu}}\left(a_--a_+^\dag+a_+-a_-^\dag\right),\\
 \hat p_1&=&-\frac{imf^{-1}\dot{\rho}}{2\sqrt{\nu}}\left(a_--a_+^\dag+a_+
 -a_-^\dag\right)-\frac{\sqrt{\nu}}{2\rho}\left(a_-+a_+^\dag+a_++a_-^\dag\right),\\
\hat  x_2&=&\frac{\rho}{2\sqrt{\nu}}\left(a_--a_+^\dag-a_++a_-^\dag\right),\\
 \hat p_2&=&\frac{mf^{-1}\dot{\rho}}{2\sqrt{\nu}}\left(a_--a_+^\dag-a_+
 +a_-^\dag\right)-i\frac{\sqrt{\nu}}{2\rho}\left(a_-+a_+^\dag-a_+-a_-^\dag\right).
\end{eqnarray}
In particulary
\begin{eqnarray}
 \hat x_1-i\hat x_2&=&\frac{i\rho}{\sqrt{\nu}}\left(a_+^\dag-a_-\right),\,\,\, \hat x_1+i\hat x_2=\frac{i\rho}{\sqrt{\nu}}\left(a_-^\dag-a_+\right),\\
\hat p_1+i\hat p_2&=&\frac{imf^{-1}\dot{\rho}}{\sqrt{\nu}}\left(a_-^\dag-a_+\right)
 -\frac{\sqrt{\nu}}{\rho}\left(a_-^\dag+a_+\right), \\
 \hat p_1-i\hat p_2&=&\frac{imf^{-1}\dot{\rho}}{\sqrt{\nu}}\left(a_+^\dag-a_-\right)
 -\frac{\sqrt{\nu}}{\rho}\left(a_+^\dag+a_-\right).
\end{eqnarray}

 \section*{Acknowledgments} 
 
 L. M. Lawson acknowledges  the  receipt of  the grant from the 
Abdus Salam international Centre  for Theoretical Physics (ICTP) Trieste
Italy and from the  German  Academic  Exchange  Service  (DAAD)

\end{document}